\documentclass{aa}  

\usepackage{graphicx}
\usepackage{txfonts}
\begin{document}

   \title{Characterisation of the new target of the NASA Lucy mission: asteroid 152830 Dinkinesh (1999 VD57)}
   \titlerunning{Characterisation of 1999 VD57}

   \author{J. de Le\'{o}n
          \inst{1,2}
          \and J. Licandro\inst{1,2}
          \and N. Pinilla-Alonso\inst{3}
          \and N. Moskovitz\inst{4}
          \and T. Kareta\inst{4}
          \and M. Popescu\inst{5}
          }

   \institute{
   Instituto de Astrof\'{i}sica de Canarias (IAC), C/V\'{i}a L\'{a}ctea s/n, E-38205 La Laguna, Tenerife, Spain\\
   \email{jmlc@iac.es}
   \and Departamento de Astrof\'{i}sica, Universidad de La Laguna, Tenerife, E-38206 La Laguna Tenerife, Spain\\
   \and Florida Space Institute -- University of Central Florida, 12354 Research Parkway, Orlando, FL 32826-0650, USA\\
   \and Lowell Observatory, 1400 W Mars Hill Road, Flagstaff, AZ 86001, USA\\
   \and Astronomical Institute of the Romanian Academy, 5 Cu\c{t}itul de Argint, 04557 Bucharest, Romania
             }

   \date{Received XXX X 2023; accepted March 8, 2023}

 
  \abstract
   {The NASA Lucy mission is aimed at the study of the very interesting population of Jupiter Trojans, considered as time capsules from the origin of our solar system. During its journey, the mission will pass near a main belt asteroid, Donaldjohanson. Recently, NASA has announced that a new asteroid in the belt will also be visited by Lucy: 152830 Dinkinesh (1999 VD57).}
   {The main goal of this work is to characterise this newly selected target, asteroid Dinkinesh, in order to provide critical information to the mission team. This information includes asteroid's most likely surface composition, albedo, and size, that will be used to better plan the data acquisition strategy at the time of the fly-by.}
   {To achieve these goals we have obtained visible spectra, colour photometry, and time-series photometry of Dinkinesh, using several telescopes located at different observatories. For the spectra we used the 10.4m Gran Telescopio Canarias (GTC), in the island of La Palma (Spain); for the colour photometry the 4.3m Lowell Discovery Telescope (LDT), near Happy Jack, Arizona (USA) was used; and for the time-series photometry we used the 82cm IAC80 telescope located in the island of Tenerife (Spain). The obtained visible reflectance spectrum was used to get the asteroid taxonomical class and so, to constrain its albedo value. Colour and time-series photometry were used to compute the asteroid absolute magnitude and, together with the albedo estimation, to constrain its size.}
   {Both visible spectrum and reflectance values computed from colour photometry show that Dinkinesh is an S-type asteroid, i.e., it is composed mainly of silicates and some metal. According to observations done by the NEOWISE survey, S-type asteroids have typical geometric albedo of $p_V$ = 0.223 $\pm$ 0.073. From our time-series photometry, we obtained an asteroid mean magnitude $r'$ = 19.99 $\pm$ 0.05, which provides an absolute magnitude $H_{r'}$ = 17.53 $\pm$ 0.07 assuming $G$ = 0.19 $\pm$ 0.25 for S-types. Using our colour-photometry, we transformed $H_{r'}$ to $H_V$ = 17.48 $\pm$ 0.05. This value of absolute magnitude combined with the geometric albedo provides a mean diameter for Dinkinesh of $\sim$900 m, ranging between a minimum size of 542 m and a maximum size of 1309 m.}
   {}

   \keywords{Minor planets, asteroids: individual: Dinkinesh, 1999 VD57 --
                methods: observational --
                techniques: spectroscopic, photometric
               }

   \maketitle
%

\section{Introduction}
   
The NASA Lucy mission, launched in October 2021, will perform the first spacecraft exploration of the Jupiter Trojans. This group of objects resides in the very stable L4 and L5 Lagrangian points of the orbit of Jupiter and are estimated to be nearly as populous as the main belt, but far more homogeneous, with low albedos and a colour bi-modality in the visible to near-infrared wavelengths \citep{Emery2015}. According to the most recent dynamical models, like the "jumping-Jupiter" version of the Nice model \citep{Walsh2011}, Trojans are considered remnants from the primordial trans-Neptunian belt and so, contain pristine information about the origins of our Solar system. For more information about the Trojans and the Lucy mission see \citet{Levison2021} and references therein. 

Lucy will accomplish its mission with a series of targeted close flybys of seven Trojans: (3548) Eurybates and its small satellite Queta, (15094) Polymele, (11351) Leucus, and (21900) Orus, all in the L4 (leading) Trojan cloud, and the Patroclus-Menoetius binary system, in the L5 Trojan cloud. It will also visit one main-belt asteroid, (52246) Donaldjohanson, in the Erigone collisional family \citep{Souza-Feliciano2020}. When launched, Lucy was scheduled to get the first close-up view of a target in 2025, but recently, a new target in the main-belt has been added to this list. After a series of small manoeuvres that will start in May 2023, the spacecraft will make an approach to this new target at a distance of about 450 km on November 1, 2023.

The selected target is the asteroid 152830 Dinkinesh (1997 VD57, Dinkinesh hereafter). It was discovered in 1999 by the LINEAR survey\footnote{\url{https://www.minorplanetcenter.net/db\_search/show\_object?object\_id=152830}}, and recently identified as a potential fly-by target for the Lucy mission by researchers at the Nice Observatory, in France. Dinkinesh is located in the inner part of the belt, having proper orbital elements $a_p$=2.19137 au, $e_p$=0.1469, and sin($i_p$)=0.0215. According to the AstDyS-2 webpage\footnote{\url{https://newton.spacedys.com/astdys/index.php?pc=0}} it is considered a background asteroid, i.e., it is associated to no collisional family. Apart from its dynamical properties, little was known about this object, with the only exception of its absolute magnitude, $H$=17.4 according to the Minor Planet Center. Therefore, the team contacted all the mission collaborators with access to telescope facilities and observational time, in an attempt to better characterise this new target. In this work we present the results from the observations done by researchers of the Solar System Group at the Instituto de Astrof\'{i}sica de Canarias, using the telescope facilities of the Observatorios de Canarias (OOCC), in collaboration with researchers from the Lowell Observatory in Arizona (USA), using their Lowell Discovery Telescope. We describe the observations and the data reduction process in Sect. \ref{sec:obs}. Results and discussion are presented in Sect. \ref{sec:results} and we present the conclusions in Sect. \ref{sec:conclusions}. 


\section{Observations and Data Reduction}
\label{sec:obs}

\subsection{Spectroscopy}

Visible spectrum of Dinkinesh was obtained on the night of December 2, 2022, using the OSIRIS camera-spectrograph \citep{Cepa2000,Cepa2010} at the 10.4m Gran Telescopio Canarias (GTC), under the program GTC32-22B. The GTC is located at the El Roque de Los Muchachos Observatory (La Palma, Canary Islands, Spain). The OSIRIS instrument is equipped with two 2K x 4K pixels detectors and a total unvignetted field of view of 7.8 x 7.8 arcmin$^2$. 
We used the 1.2" slit and the R300R grism (dispersion of 7.74 \AA/pixel, resolution $R$ = 348 for a 0.6" slit) covering the 0.48 - 0.92 $\mu$m wavelength range. The slit was oriented along the parallactic angle to minimise the effects of atmospheric differential refraction and the telescope tracking was set at the asteroid's proper motion. Details on the observational circumstances are shown in Table \ref{tab:obs}. Two spectra of 900 s of exposure time each were obtained, with an offset of 10" in the slit direction in between them. To obtain the asteroid's reflectance spectrum, we observed solar analogue stars from the Landolt catalogue \cite{Landolt1992} SA93-101, SA98-978, and SA102-1081 at a similar airmass as that of the asteroid. 

\begin{figure*}[h!]
    \begin{center}
    \includegraphics[width=1.3\columnwidth]{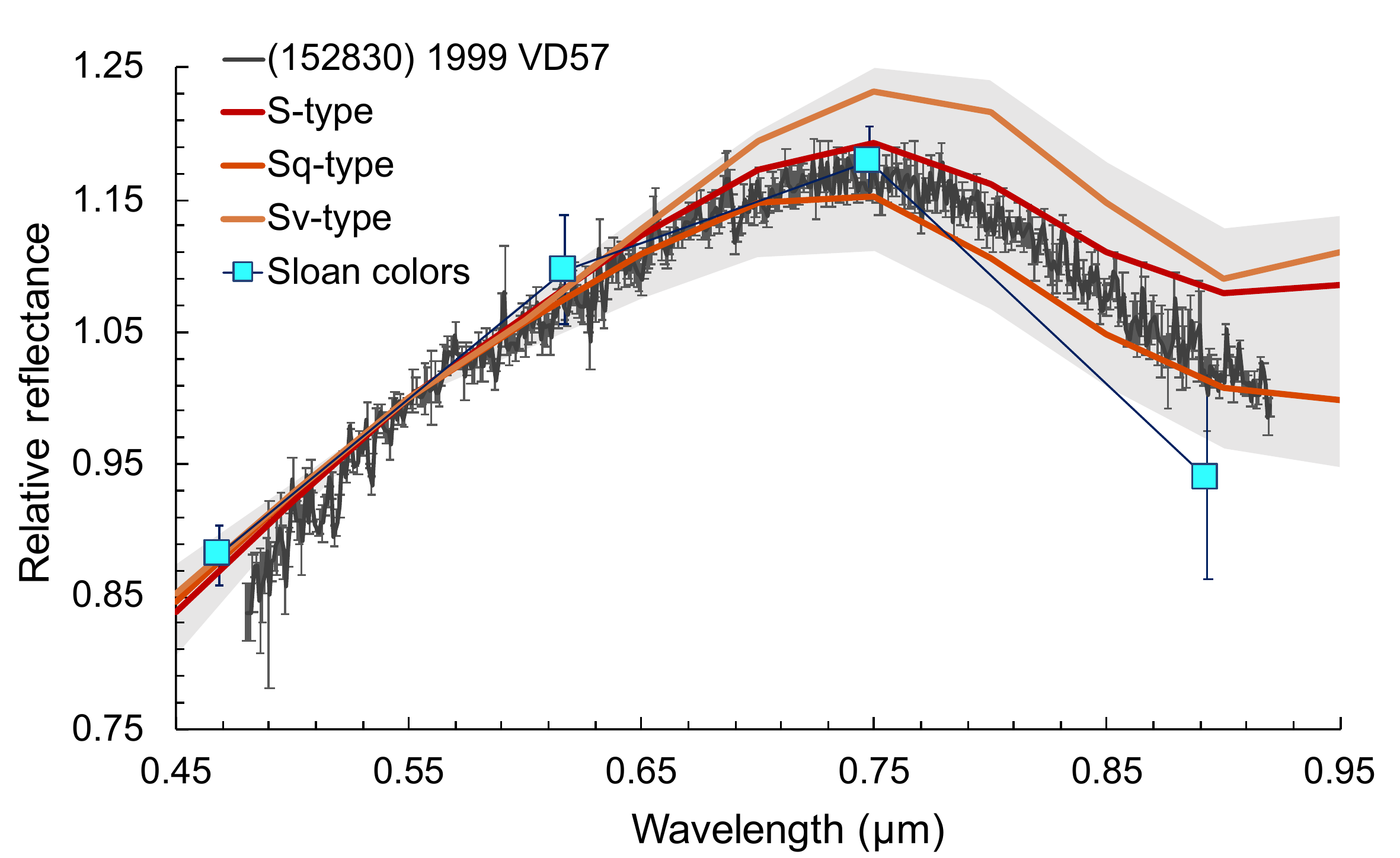}
    \caption{Visible spectrum of asteroid Dinkinesh obtained with the 10.4m Gran Telescopio Canarias (GTC). Error bars correspond to the 1$\sigma$ of the average of the individual spectra. Reflectance values from the colour photometry obtained with the 4.3m Lowell Discovery Telescope (LDT) using Sloan $griz$ filters are shown as blue squares. Coloured lines are the template spectra of the S-type (red), Sq-type (orange), and Sv-type (yellow) taxons from the \citet{DeMeo2009} taxonomy. The hatched region accounts for the $\pm1\sigma$ dispersion of the spectral classes.}
    \label{fig:VD57spectrum}
    \end{center}
\end{figure*}

Data reduction was done using standard procedures (see \citealt{deLeon2016}). The images were initially bias and flat-field corrected. Sky background was subtracted and a one-dimensional spectrum was extracted using a variable aperture, corresponding to the pixel where the intensity was 10\% of the peak value. Wavelength calibration was applied using Xe+Ne+HgAr lamps. This procedure was applied to the spectra of the asteroid and the stars. We then divided the asteroid's individual spectra by the spectra of the solar analogues, and the resulting ratios were averaged to obtain the final reflectance spectrum of Dinkinesh, shown in Fig. \ref{fig:VD57spectrum}. 

\begin{table}
\caption{Observational circumstances of Dinkinesh. Observation type includes time-series photometry (Phot.), colour photometry (Colo.) and visible spectra (Spec.).}
\label{tab:obs}
\centering
\begin{tabular}{cccccc}
\hline\hline\\[-3mm]
Obs. & Date & $m_V$ & $\alpha$ & $\Delta$ & $r$ \\
Type & Obs. (UTC) & & ($^{\circ}$) & (au) & (au) \\
\hline\\[-3mm]
Phot. & 2022 11 17.0009 & 20.3 & 23.0 & 1.161 & 1.947\\
Colo. & 2022 11 19.3815 & 20.2 & 22.2 & 1.142 & 1.947\\
Spec. & 2022 12 02.0729 & 19.8 & 16.8 & 1.052 & 1.946\\
\hline
\end{tabular}
\end{table}

\subsection{Time-series photometry}
\label{lightcurve}

Time series photometry of Dinkinesh was obtained on November 17, 2022 using the IAC80 telescope at Teide Observatory (Tenerife, Canary Islands, Spain). For the observational circumstances see the information in Table \ref{tab:obs}. The IAC80 is a 82cm telescope with  $f/D =$ 11.3 in the Cassegrain focus configuration. It is equipped with the CAMELOT-2 camera, with a back-illuminated e2v 4K x 4K pixels CCD of 15 $\mu$m$^2$ pixels, a plate scale of 0.32 arcsec/pixel and a field of view of 21.98 x 22.06 arcmin$^2$. We used a Sloan $r$ filter and 70 s exposure time with the telescope in sidereal tracking.

\begin{figure}[h!]
    \begin{center}
    \includegraphics[width=\columnwidth]{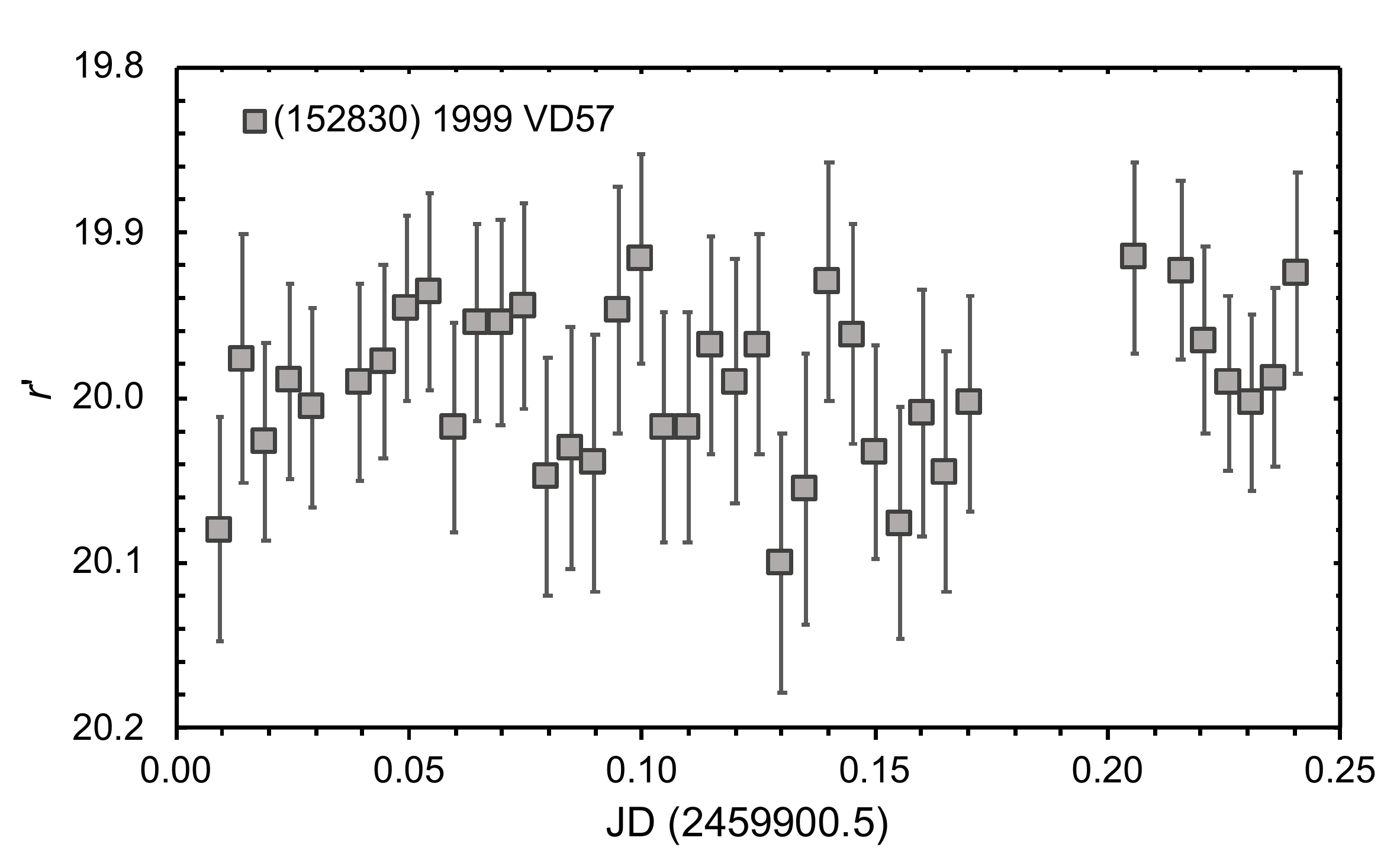}
    \caption{Light curve of Dinkinesh obtained on November 17, 2022 with the 80cm IAC80 telescope.}
    \label{fig:lightcurve}
    \end{center}
\end{figure}

Images were bias and flat-field corrected using standard procedures. Aperture photometry was done using the Photometry Pipeline\footnote{\url{https://photometrypipeline.readthedocs.io/en/latest/}} (PP) software \citep{Mommert2017}. PP uses the Source Extractor software for source identification and aperture photometry and the SCAMP\footnote{\url{https://www.astromatic.net/software/scamp/}} software for image registration. Image registration and photometric calibration are based on matching field stars with the USNO-B1 \citep{Monet2003} and PanSTARRS catalogues, respectively. The circular aperture photometry is performed using Source Extractor; an optimum aperture radius is identified using a curve-of-growth analysis by the PP. The final calibrated photometry for each field source is written into a queryable database, and target photometric results are extracted from this database. Moving targets are identified using JPL Horizons ephemerides service \citep{Giorgini1996}. The resulting light curve is shown in Fig\ref{fig:lightcurve}. The observed gap between Julian Dates JD 2459900.67 and JD 2459900.70 is because the measured asteroid magnitude was affected by a close and bright star.

\subsection{Colour photometry}

Broad-band colours were obtained with the Large Monolithic Imager (LMI) at the 4.3m Lowell Discovery Telescope (LDT) on UT 19 November 2022. LMI is a 6k x 6k e2v CCD that images a 12.3 arcmin$^2$ field of view. The images were obtained using the Sloan Digital Sky Survey (SDSS) $griz$ filter set and were sampled at 0.36 arcseconds per pixel in 3x3 binning mode. Exposures were 60 seconds each, with the filters sequenced in an interleaved pattern, e.g. $r-g-r-i-r-z-r$. Given the unknown rotational lightcurve of Dinkinesh, this strategy provided a means to correct for rotational variability from one exposure to the next by monitoring the magnitude of the target in $r$-band. This can be important for deriving colours (independent of rotation effects) when the individual filters are collected non-simultaneously. However, in this case, no significant lightcurve variability was detected over a $\sim$33 minute observing window, consistent with the time-series photometry presented in Sect. \ref{lightcurve}. A total of 4, 5, 4, and 8 exposures were taken in $g$, $r$, $i$, and $z$ bands respectively. The extra z-band exposures were taken to offset the lower signal-to-noise per exposure in that band. Observing conditions were sub-optimal with poor seeing ($\sim$3") and significant cloud cover, which caused variability at the level of 0.5 - 1 magnitude in the zero points for each filter.

Reduction of the images followed standard bias and flat field correction techniques. The photometry was also measured with the Photometry Pipeline, and the photometric calibration of each frame based on PanSTARRS field stars served to counteract the effects of the variable extinction. For each filter a weighted mean magnitude was computed and the differences of these magnitudes were used to derive colours. The weighted mean magnitudes and average errors are $\bar{g}=20.57\pm0.03$, $\bar{r}=19.90\pm0.03$, $\bar{i}=19.71\pm0.03$, and $\bar{z}=19.95\pm0.10$. This yields the following colours: ($g-r$) = 0.67 $\pm$ 0.04, ($r-i)$ = 0.19 $\pm$ 0.04, and ($r-z$) = -0.05 $\pm$ 0.10. These colours are transformed into flux or reflectance values using the SDSS estimate for solar magnitudes \citep{Holmberg2006}: $g_\odot=5.12$, $r_\odot=4.68$, $i_\odot=4.57$, and $z_\odot=4.54$. The resulting reflectance values are plotted in Fig.\ref{fig:VD57spectrum} as blue squares. For comparison purposes, the reflectance values are represented with a normalisation at 0.55 $\mu$m, which is calculated based on a linear interpolation of the data between the $g$ and the $r$ bands.

\section{Results and Discussion}
\label{sec:results}

Visible spectroscopy enabled us to perform the taxonomical classification of the asteroid. We did so by using the M4AST on-line tool\footnote{\url{http://spectre.imcce.fr/m4ast/index.php/index/home}} \citep{Popescu2012}, which fits a curve to the data and compare it to the taxons defined by \citealt{DeMeo2009} using a $\chi^2$ fitting procedure. A list with the best three results in order of decreasing goodness of fit is provided. For the case of Dinkinesh, the best three fits are S, Sq, and Sv (see Fig. \ref{fig:VD57spectrum}). The reflectance values obtained from the visible colours are in very good agreement with this classification. Therefore we can confidently conclude that the asteroid belongs to the S-complex. 

The taxonomical classification of Dinkinesh provides a statistical estimation of its albedo and consequently, a more solid estimation of its size. Using measurements of the NEOWISE survey, \citet{Mainzer2011} obtained a median albedo value of $p_V$ = 0.223, with a maximum value of 0.557 and a minimum value of 0.114 for the \citet{DeMeo2009} S-complex. 

The computed asteroid mean magnitude in the 6 h period of observations with the IAC80 on November 17 was $\bar{r}=19.99 \pm 0.05$. Due to the asteroid's apparent visual magnitude at the time of observations (see Table \ref{tab:obs} and the IAC80 telescope's small aperture, the measured signal-to-noise ratio was only $\sim$10. Nevertheless, it allowed us to conclude that the asteroid brightness does not vary more than 0.1 mag in a 6 h period. On the other hand, during the observations done with the LDT on November 19 the asteroid's mean magnitude was $\bar{r}=19.90\pm0.03$. 

Assuming a $G$-value for the S-complex taxonomy, we can derive the asteroid's absolute magnitude using the $H-G$ phase function described in \cite{Muinonen2010}. It has been commonly assumed a slope of $G = 0.15$ for asteroids, but several works show that this value can be different for different taxonomical classes. Recently, \citet{Colazo2021} used photometry of asteroids from the Gaia DR2 and the Asteroid Photometric Catalogue \citep{Lagerkvist1995} to obtain mean phase-slopes for S-, C-, D-, and X-type objects. In their Table 3, they compare the obtained results with those obtained from PanSTARRS PS1 \citep{Veres2015} and those from \citet{Pravec2012} using their own observations and data from \citet{Warner2009}. In the case of S-type asteroids, Colazo et al. obtained a value of $G = 0.19 \pm 0.44$, while Vere\v{s} et al. obtained $G = 0.16 \pm 0.26$ and Pravec et al. $G = 0.23 \pm 0.05$. We computed a simple average of these three values, propagating their associated errors, to obtain $G = 0.19 \pm 0.25$. Using this value we obtain an absolute magnitude of $H_r = 17.13 \pm 0.07$ on November 17 and $H_r = 17.15 \pm 0.04$ on November 19 (notice that both values agree well within the uncertainties). To compute $H_V$ we used our colour photometry and the transformation equations from \citet{Jesser2005}. Our final absolute magnitude for Dinkinesh is $H_V = 17.48 \pm 0.05$. 

\begin{figure}[h!]
    \begin{center}
    \includegraphics[width=\columnwidth]{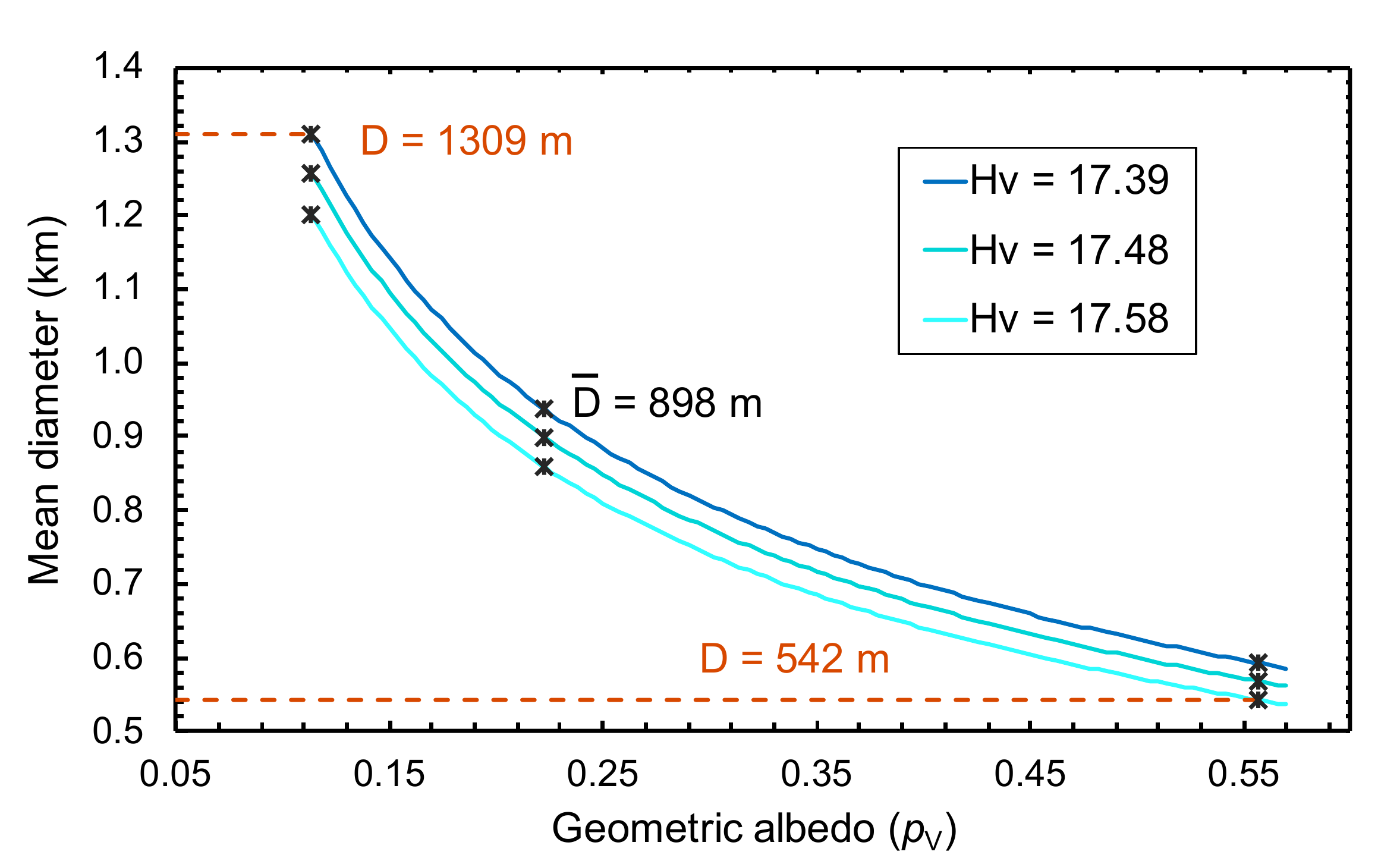}
    \caption{Dinkinesh's mean estimated diameter (898 m) and values computed for the range of albedo values associated to its taxonomical class, S-complex (black crosses), i.e., 0.114 < $p_V$ < 0.557 \citep{Mainzer2011}, the range of $G$ values for S-types, i.e., 0.16 < $G$ < 0.23, and the corresponding range in $H_V$, using those $G$ values, that goes from $H_V$ = 17.39 to $H_V$ = 17.58.}
    \label{fig:DvsAlbedo}
    \end{center}
\end{figure}
 
Using this value and the geometric albedo associated to S-types, $p_V=0.223$, we estimate a mean diameter for Dinkinesh, using the standard diameter-albedo relationship $\bar{D}$ = 10$^{(-0.2H)}$1329/$\sqrt{p_V}$ = 0.898 km. Considering the range in $G$ values 0.16 < $G$ < 0.23, and the corresponding range in computed $H_V$, 17.44 $\pm$ 0.05 < $H_V$ < 17.53 $\pm$ 0.05, as well as and the minimum and maximum values for the geometric albedo (see Fig. \ref{fig:DvsAlbedo}), the estimated mean diameter can vary from 542 m to 1309 m. The upper limit of the amplitude of the lightcurve (0.1 mag in a 6 h period) suggests that either the shape of Dinkinesh is not very elongated, that the polar axis was oriented close to the line-of-sight during the observations, or that the asteroid is a very slow rotator. 

\section{Conclusions}
\label{sec:conclusions}

We presented here visible spectra and colours, as well as visible time-series photometry of main belt asteroid 152830 Dinkinesh (1999 VD57), recently selected by NASA Lucy mission as an additional target in its journey to Jupiter Trojans. Dinkinesh's visible spectrum and time-series photometry were obtained with the 10.4m Gran Telescopio Canarias and the 82cm IAC80 telescopes, respectively (Spain), while visible colour photometry was obtained with the 4.3m Lowell Discovery Telescope (USA). Our conclusions can be summarised as follows.
   \begin{enumerate}
      \item From its visible spectrum and colours, Dinkinesh can be classified as an S-type asteroid, i.e., it is mainly composed of silicates and metal.
      \item From colour and time-series photometry we compute an absolute magnitude $H_V$=17.48 $\pm$ 0.05, in good agreement with the value reported by the Minor Planet Center ($H_V$=17.4). Observations of Dinkinesh at different phase angles will enable a more accurate determination of its absolute magnitude, hence, a better estimation of its mean diameter.
      \item Using the geometric albedo value associated to S-type asteroids ($p_V$=0.223) we obtain that Dinkinesh has mean diameter of $\bar{D}$=898 m, with a range of values that goes from $D$=542 m to $D$=1309 m. 
      \item The obtained upper limit of the lightcurve amplitude (0.1 mag in a 6 h period) is not conclusive, and only allows to suggest several possibilities, including the asteroid being a slow rotator. 
   \end{enumerate}

\begin{acknowledgements}
      JdL and JL acknowledge support from the ACIISI, Consejer\'{i}a de Econom\'{i}a, Conocimiento y Empleo del Gobierno de Canarias and the European Regional Development Fund (ERDF) under grant with reference ProID2021010134. NPA acknowledges support from the Center for Lunar and Asteroid Surface Science (CLASS) a NASA's SSERVI team funded in CAN3. The work of MP is financed by a grant of the Romanian National Authority for Scientific Research and Innovation CNCS -- UEFISCDI, project number PN-III-P2-2.1-PED-2021-3625. Based on observations made with the Gran Telescopio Canarias (GTC), installed at the Spanish Observatorio del Roque de los Muchachos of the Instituto de Astrof\'{i}sica de Canarias, on the island of La Palma.
      
\end{acknowledgements}

\bibliographystyle{aa}      
\bibliography{biblioLucy}   

\end{document}